\documentclass[namedreferences]{solarphysics}
\usepackage{subfig}
\usepackage{savesym}
\savesymbol{iint}
\savesymbol{iiint}
\usepackage{wasysym}
\usepackage{mathtools}
\usepackage{xcolor}
\usepackage{amsmath}
\usepackage{mathrsfs}
\usepackage{picture}
\usepackage[hyperref,optionalrh,solaromanenum]{spr-sola-addons} 
\usepackage{graphicx}                    
\usepackage{epstopdf}
\usepackage{amssymb,bm}                    
\usepackage{color}                       
\usepackage{breakurl}                         
\usepackage{hhline} 
\usepackage{booktabs}
\usepackage{dirtytalk}

\newcommand{\br}{\bm{r}}

\newcommand{\ii}{\textrm{i}}

\newcommand{\Nobs}{N}
\newcommand{\wn}{\omega_{\rm max}}

\chardef\us=`\_

\usepackage{tikz}

\begin{document}

\newcommand{\AB}[1]{\textcolor{red}{[AB: #1]}}
\newcommand{\DF}[1]{\textcolor{blue}{[DF: #1]}}
\newcommand{\CSH}[1]{\textcolor{green}{[CSH: #1]}}
\newcommand{\DY}{\textbf}

\begin{article}

\begin{opening}

\title{Comparison of Travel-Time and Amplitude Measurements for Deep-Focusing Time--Distance Helioseismology}
\author[addressref=aff1,corref,email={pourabdian@mps.mpg.de}]{\inits{M.}\fnm{Majid}~\lnm{Pourabdian}}
\author[addressref=aff1]{\inits{D.}\fnm{Damien}~\lnm{Fournier}}
\author[addressref={aff1,aff2,aff3}]{\inits{L.}\fnm{Laurent}~\lnm{Gizon}}

\address[id=aff1]{Max-Planck-Institut f\"ur Sonnensystemforschung, 37077 G\"ottingen, Germany}
\address[id=aff2]{Institut f\"ur Astrophysik, Georg-August-Universit\"at G\"ottingen, 37077 G\"ottingen, Germany}
\address[id=aff3]{Center for Space Science, NYUAD Institute, New York University Abu~Dhabi,  Abu~Dhabi, UAE}

\runningauthor{M. Pourabdian \it et al.}
\runningtitle{Travel-Time and Amplitude Measurements for Deep-Focusing Helioseismology}

\begin{abstract}
The purpose of deep-focusing time--distance helioseismology is to construct seismic measurements that have a high sensitivity to the physical conditions at a desired target point in the solar interior.  With this technique, pairs of points on the solar surface are chosen such that acoustic ray paths intersect at this target (focus) point. 
Considering acoustic waves in a homogeneous medium, we compare travel-time and amplitude measurements extracted from the deep-focusing cross-covariance functions.
Using a single-scattering approximation, we find that the spatial sensitivity of deep-focusing travel times to sound-speed perturbations  is zero at the target location and maximum in a surrounding shell. This is unlike the deep-focusing amplitude measurements, which have maximum sensitivity at the target point. We compare the signal-to-noise ratio for travel-time and amplitude measurements for different types of sound-speed perturbations, under the assumption that noise is solely due to the random excitation of the waves. We find that, for highly localized perturbations in sound speed, the signal-to-noise ratio is higher for amplitude measurements than for travel-time measurements. We conclude that amplitude measurements are a useful complement to travel-time measurements in time--distance helioseismology. 
\end{abstract}
       
\end{opening}

\section{Introduction} 
Time--distance helioseismology \citep[][]{Duvall_etal_1993} is a branch of local helioseismology  \citep[\emph{e.g.}][]{Gizon_Birch_2005} that aims at probing the complex subsurface structures of the solar interior. The time--distance method measures the travel times of acoustic waves between any pair of points on the solar surface from the cross-covariance function of the observed oscillation signals. Seismic travel times contain information about the local physical properties of the medium and have thus been broadly used in helioseismology \citep[\emph{e.g.}][]{Gizon_Birch_2002, Birch_etal_2004, Gizon_etal_2010}.

  
A consistent issue with local helioseismology is the signal-to-noise ratio. When examining near-surface structures such as supergranular flows \citep[][]{Duvall_etal_1996,Langfellner_etal_2015}, averaging is typically performed around an annulus, where the cross-covariance is calculated between the center point and the average signal in the annulus. This technique is highly sensitive to near-surface perturbations.
To probe greater depths, one would seek a different averaging technique that has peak sensitivity at any chosen target depth. Such a technique is known as deep-focusing and was first described by \citet{Duvall_1995}, who outlined a procedure in which points on the surface are chosen such that a large number of connecting ray paths intersect at the target (focus) point, with the  expectation that sensitivity is large near the target depth.
The deep-focusing time--distance technique has been employed to study the meridional flow in the solar interior \citep[\emph{e.g.}][]{Hartlep_etal_2013,Zhao_etal_2013} and  sunspot structure \citep[\emph{e.g.}][]{Moradi_Hanasoge_2010}. 
 \citet{Jensen_2001} investigated the application of the deep-focusing method to improve inversions for large sunspots. Using the Rytov approximation, he found sensitivity in a shell around the target point but zero sensitivity at the target point, consistent with wavefront healing seen in the Born approximation in geophysics and helioseismology. To resolve this drawback, \citet{Hughes_etal_2007} suggested an optimized technique for deep focusing that allocates weightings for each measurement. They obtained improvements in the results by considering travel-time measurements of synthetic experiments. 
 
In addition to the travel times, the cross-covariance function contains additional information that may be of use to helioseismology. For instance, in terrestrial seismology cross-covariance amplitudes have been used to characterize seismic waves \citep[\emph{e.g.}][]{Nolet_etal_2005}. The importance of the amplitudes was examined by \citet{Dalton_Hjorleifsdottir_2014}, who concluded that assumptions and simplifications in the measurement of surface-wave amplitudes affect the attenuation structure found through inversions. Moreover, \citet{Dahlen_Baig_2002} investigated the Fr\'echet sensitivity kernels using the geometrical ray approximation for travel-time and amplitude measurements. They found a maximum sensitivity along the point-to-point ray path when examining the amplitude of seismic-wave cross-correlation.  
In contrast to travel times, few studies have considered the amplitude measurements of the cross-covariance function in helioseismology. \citet{Liang_etal_2013} measured the spatial maps of wave travel times and amplitudes from the cross-covariance function of the wave field around a sunspot in the NOAO Active Region 9787. Using 2D ray theory, they observed an amplitude reduction that was attributed to the defocusing of wave energy by the fast-wave-speed perturbation in the sunspot. Recent work by \citet{Nagashima_etal_2017} described a linear procedure to measure the amplitude of the cross-covariance function of solar oscillations. This linear relation between the cross-covariance function and the amplitude allows the derivation of Born sensitivity kernels using the procedure of \citet{Gizon_Birch_2002}, which provides a straightforward interpretation for the amplitude measurements.

The deep-focusing time--distance technique using amplitude measurements is lacking in time--distance helioseismology. Furthermore, the deep-focusing analysis has been considered only using the ray theory, which is a high-frequency approximation and does not take into account finite-wavelength effects. As a result, the ray approximation may be inaccurate for amplitude calculations \citep[\emph{e.g.}][]{Tong_etal_1998}. In this study, we use the deep-focusing time--distance technique to compare signal and noise for travel-time and amplitude measurements under the Born approximation. 
Section \ref{sec:cross-covariance} describes the definition of travel-time and amplitude measurements and explains the deep-focusing technique and the noise model. The setup and derivation of sensitivity kernels are explained in Section \ref{sec:toy_problem} and the results are presented in Section \ref{sec:results}. Conclusions are given in  Section \ref{sec:summary}.


\section{Travel-Time and Amplitude Measurements}
\label{sec:cross-covariance}

\subsection{Definitions}

In time--distance helioseismology, one uses the cross-covariance function between the oscillation signals observed at any two points [$\bm{r}_1$ and $\bm{r}_2$] on the solar surface to recover the desired information within the relevant wave-field observable. In general, we observe the line-of-sight velocity $[\phi]$  and define the temporal cross-covariance function for surface locations $\bm{r}_1$ and $\bm{r}_2$ as    
\begin{equation}
 C(\bm{r}_1,\bm{r}_2,t) = \frac{1}{T} \int_{-T/2}^{T/2} \phi(\bm{r}_1,t')\phi(\bm{r}_2,t'+t) \, \mathrm{d} t',  
\label{eq:cross-cov_general}
\end{equation}
where $t$ is the time lag and $T$ is the duration of observation. Considering small changes to a reference solar model, one can define the incremental travel time $[\delta\tau]$ and relative amplitude $[\delta a]$ between the observed $[C(\bm{r}_1,\bm{r}_2;t)]$ and reference $[C_0(\bm{r}_1,\bm{r}_2;t)]$ cross-covariances as  
\begin{eqnarray}
 \delta \tau(\bm{r}_1,\bm{r}_2) &=& \int_{-T/2}^{T/2}  W_\tau(\bm{r}_1,\bm{r}_2,t)  \, \delta   C(\bm{r}_1,\bm{r}_2,t)    \, \mathrm{d} t, \\
 \delta a (\bm{r}_1,\bm{r}_2) &=& \int_{-T/2}^{T/2}  W_a(\bm{r}_1,\bm{r}_2,t)  \, \delta   C(\bm{r}_1,\bm{r}_2,t)   \, \mathrm{d} t ,
 \end{eqnarray}
where 
\begin{equation}
\delta C (\bm{r}_1,\bm{r}_2,t) =  C(\bm{r}_1,\bm{r}_2,t) - C_0(\bm{r}_1,\bm{r}_2,t) .
\end{equation} 
The above linear relations between the measurements and the cross-covariance function are specified via the weighting functions $[W]$ given by \citet{Nagashima_etal_2017}: 
\begin{eqnarray}
W_{\tau}(t) &=&  -\frac{w(t)\partial_{t} C_{0}(t)}{\int_{-T/2}^{T/2}w(t')[\partial_{t'} C_{0}(t')]^2\mathrm{d}t'},  \label{eq:weighting_func} \\
 W_{a}(t) &=&  \frac{w(t)C_{0}(t)}{\int_{-T/2}^{T/2}w(t')[ C_{0}(t')]^2\mathrm{d}t'},  \label{eq:weighting_func_amp}
\end{eqnarray}
where $w(t)$ is a window function that may select the first-arrival wave packet. With this definition of the weighting function $[W_a]$ the relative amplitude $[\delta a]$ is dimensionless.

Throughout this article, we use $q$ to denote either the travel-time $[\tau]$ or the amplitude measurement $[a]$. Using this compact notation, we write 
\begin{equation}
 \delta q(\bm{r}_1,\bm{r}_2) = \int_{-T/2}^{T/2}  W_q(\bm{r}_1,\bm{r}_2,t) \delta C(\bm{r}_1,\bm{r}_2,t) \, \mathrm{d} t, 
 \quad \text{ for } q\in \{ \tau, a \} .
\label{eq:travel_time_general}
\end{equation} 

\subsection{Deep-Focusing Averages}
\label{sec:Deep_Foc}
\begin{figure}[t]
\begin{center}
\includegraphics[width=0.8\textwidth, trim= {0 0 0 {-0.05\textwidth}}]{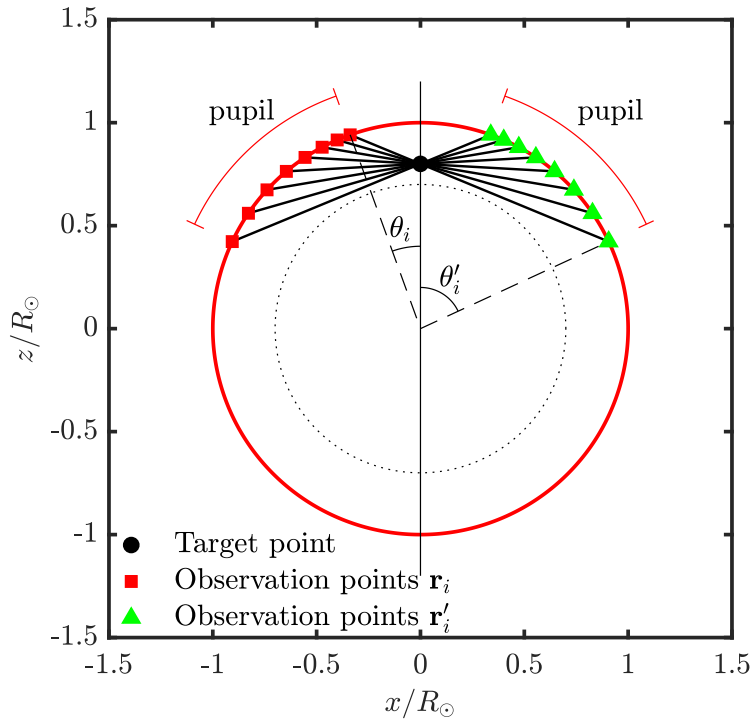}
\caption{Sketch depicting the location of the observation points $\br_{i}$ ({\it red squares}) and $\br'_{i}$ ({\it green triangles})  inside the  pupil. The points are chosen such that the ray paths ({\it black lines}) intersect at a focus point at $z_t=0.8\,\mathrm{R_\odot}$ ({\it black dot}). The {\it dotted circle} has a radius of $0.7\,\mathrm{R_\odot}$.}
\label{fig:points}
\end{center}
\end{figure}

The basic idea of the deep-focusing technique is to obtain high sensitivity to a physical quantity by focusing at a given target point. To do so, we consider a set of pairs of points on the solar surface such that the ray paths (straight lines for a homogeneous medium) intersect at a chosen target point. As an example, Figure~\ref{fig:points} illustrates how these pairs of points could be distributed on the surface of the near-side of the Sun. In a solar case, the ray paths would be curved due to the sound-speed stratification.

For any desired target point $[\br_t]$ in the solar interior, we define the averaged travel-time and amplitude perturbations as  
\begin{equation}
\overline{\delta q}(\br_t)  = \frac{1}{\Nobs} \sum_{i=1}^{\Nobs} \delta q_{i}(\br_t), \quad q \in \{\tau, a\} ,
\label{eq:ave_travel_time}
\end{equation}
where $\delta q_i(\br_t)$ represents the point-to-point measurement between the points $\br_i$ and $\br'_i$ chosen such that the ray path intersects at the focus point $\br_t$,
\begin{equation}
\delta q_{i}(\br_t)  = \delta q(\br_{i}, \br'_{i}).
\label{eq:travel_time_ij}
\end{equation} 
The observations points $\br_{i}$ and $\br_{i}'$ are on a sphere of radius $\mathrm{R_\odot}$ and have coordinates $\br_{i}=(\mathrm{R_\odot},\theta_i,\phi_i)$ and $\br'_{i}=(\mathrm{R_\odot},  {\theta}'_i, \phi'_i)$ in the spherical-polar coordinate system whose polar axis contains the target point (depicted in Figure~\ref{fig:points}).
 The index $i$ spans $[1, N]$, where $N=N_\theta N_\phi$ is the total number of pairs of points, with $N_\theta$ the number of colatitudes and $N_\phi$ the number of longitudes.
Each index $i$ is associated with a pair of indices $(i_\theta,i_\phi)\in [1,N_\theta]\times [1,N_\phi]$, where the first index refers to the colatitudes $\theta_{i_\theta}$ and $\theta'_{i_\theta}=\Delta-\theta_{i_\theta}$ (where $\Delta$ is the colatitude difference between the two observation points in a pair) and the second index refers to the uniformly-spaced azimuths  $\phi_{i_\phi} = {2\pi}(i_\phi-1)/N_\phi$ and $\phi_{i_\phi}'= \phi_{i_\phi}+\pi$.
The range of colatitudes  $\theta_1 \le \theta \le \theta_{N_\theta}$ defines the extent of the pupil.
Choosing a maximum value $\theta_{N_\theta}=65^\circ$, the value of $\theta_1$ then depends on the target depth. At a fixed longitude, the colatitudes of the points within the pupil are chosen such that the angle between neighboring ray paths is uniform.

\subsection{Noise Model} 
\label{sec:noise_model}
Here we describe the noise in the averaged measurements for travel time and amplitude. 
Random noise in helioseismology is due to the stochastic excitation of acoustic waves by turbulent convection. The noise model developed by \citet{Gizon_Birch_2004}
is based on the reasonable assumption that the reference wave field $[\phi_0]$ is described by a stationary Gaussian random process. 

The variance of the averaged travel-time or amplitude measurement is given by
\begin{align}
\sigma^2_{q}= \textrm{Var}\left[ \, \overline{\delta q} \, \right]
= \frac{1}{\Nobs^2} \sum_{i=1}^{\Nobs} \sum_{j=1}^{\Nobs} \textrm{Cov}[ \delta q_{i} , \delta q_{j} ] .
\label{eq:std_tau}
\end{align}   
The covariance between any two measurements [$\textrm{Cov}[ \delta q_{i} , \delta q_{j} ]$] depends on the reference cross-covariance function in the frequency domain, 
\begin{equation}
C_0(\omega)=\frac{1}{2\pi}\int_{-{T}/{2}}^{{T}/{2}}  C_0(t)e^{\ii\omega t}  \,  \mathrm{d}t ,
\label{C_0_freq}
\end{equation}
and on the weighting functions $[W_q]$. \citet{Fournier_Gizon_2014} showed that the covariance is explicitly given by 
\begin{multline}
\textrm{Cov}[ \delta q_{i} , \delta q_{j} ]=\frac{(2\pi)^3}{T}\int_{-\wn}^{\wn}  \mathrm{d}\omega W_{q}^*(\br_{i},\br'_{i},\omega)  \left[ W_{q}(\br_{j},\br'_{j},\omega) C_0(\br'_{i},\br'_{j},\omega)
\right.
\\ \left.
\times C_0(\br_{i},\br_{j},\omega) 
  + W_{q}^*(\br_{j},\br'_{j},\omega) C_0(\br_{j},\br'_{i};\omega)C_0(\br_{i},\br'_{j},\omega) \right],
\label{eq:noise_cov_tau}
\end{multline}
where $\wn = \pi/h_t$ is  the Nyquist frequency and $h_t$ is the temporal cadence. The star denotes complex conjugation. Note that the noise covariance was originally  derived for travel-time measurements, but it is easily extended to the amplitude measurements due to the linearity between $\delta a$ and $\delta C$. 

\section{Travel-Time and Amplitude Sensitivity Kernels for Sound-Speed Perturbations to a Uniform Background Medium}  
\label{sec:toy_problem}

\subsection{Wave Equation and Reference Green's Function}
We consider the wave equation at angular frequency $\omega$,
\begin{equation}
\mathcal{L} \phi(\br,\omega) = s(\br,\omega),
\end{equation}
where the wave operator is 
\begin{equation}
\mathcal{L}=\nabla^2 + k^2(\br,\omega)
\label{eq:wave_operator}
\end{equation}
and the wave number is given by
\begin{equation}
k(\br,\omega) =\frac{\omega}{c(\br)} \left(1 + \ii \gamma \right),
\end{equation}
where $c(\br)$ is the sound speed and $\gamma$ is a constant number that accounts for attenuation. The random source of excitation $s(\br,\omega)$ is assumed to be stationary, uniformly distributed and spatially uncorrelated throughout the medium. Under these conditions, the expectation value of the cross-covariance function can be related directly to the imaginary part of the Green's function in the frequency domain \citep[\emph{e.g.}][]{Gizon_etal_2017} 
\begin{equation}
\langle C(\omega) \rangle = \frac{\Pi(\omega)}{\omega} \mbox{Im} \,G(\br, \br',\omega), \label{eq:cross_cov}
\end{equation}
where the function $\Pi(\omega)$ is related to the frequency dependence of the source covariance. The angle brackets $\langle \ \rangle$ represent the expectation value of a stochastic quantity.

We consider a background medium with the reference wave number 
\begin{equation}
k_0(\omega) =\frac{\omega}{c_0} \left(1 + \ii \gamma \right),
\label{eq:ref_k}
\end{equation}
where the reference sound speed is constant $c_0$. The reference Green's function is solution of  $\mathcal{L}_0G_0(\br, \br')=\delta(\br - \br')$ where $\mathcal{L}_0=\nabla^2 + k_0^2$ is the reference wave operator. Using a Sommerfeld radiation condition to avoid incoming waves at infinity, the expression for $G_0$ is 
\begin{equation}
G_0(\br,\br')=-\frac{e^{\ii k_0\|\br - \br'\|}}{4\pi\|\br - \br'\|}.
\label{eq:analytic_Green}
\end{equation}
This simple analytic expression motivates the choice that we have made of a uniform medium.

\subsection{Perturbation to the Cross-Covariance Function}
\label{sec:Born approximation}
In this section we compute the  perturbation to the cross-covariance $[\delta C=C-C_0]$ due to a small perturbation in sound speed $[\delta c (\br) = c(\br)-c_0]$. Using Equation~\ref{eq:cross_cov}, the expectation value of $\delta C$ is related to the perturbation to the Green's function,
\begin{equation}
\langle \delta C(\omega) \rangle =  \langle C(\omega) \rangle  -  C_{0}(\omega) 
 = \frac{\Pi(\omega)}{\omega} \mbox{Im} \, (\delta G).
\label{eq:smartgreen}
\end{equation}
Under the first-order Born approximation  we have
\begin{equation}
\mathcal{L}_0\delta G(\br, \br',\omega)=-\delta \mathcal{L}G_0(\br, \br',\omega),
\label{eq:Born}
\end{equation}
where $\delta \mathcal{L} = -2k_0^2\delta c(\br)/c_0$ is the perturbation to the wave operator caused by the perturbations in the sound speed $\delta c$. According to Equation~\ref{eq:Born}, the Born approximation is an equivalent-source description of wave interaction. Using $G_0$ to solve for $\delta G$, we find
\begin{equation}
\delta G(\br_{i}, \br'_{i},\omega) = \int_V G_0(\br_{i},\bm{r},\omega)
\, 2k_0^2 \frac{\delta c(\br)}{c_0} \,  G_0(\bm{r},\br'_{i},\omega) \, \mathrm{d}^3\bm{r}, \label{eq:deltaG}
\end{equation}
where $V$ is the computational domain, including the full sphere. It follows that the perturbation to the cross-covariance between the points $\br_{i}$ and $\br'_{i}$ is
\begin{equation}
\langle \delta C_{i}(\omega) \rangle = \int_V \mathscr{C}_{i}(\br,\omega) \frac{\delta c(\br)}{c_0} \,  \mathrm{d}^3\bm{r}, \label{eq:deltaC}
\end{equation}          
where $\mathscr{C}_{i}(\br,\omega)$ is defined as
\begin{equation}
\mathscr{C}_{i}(\br,\omega) = \frac{2\Pi(\omega)}{\omega} \mbox{ Im} \left[ k_0^2G_0(\bm{r},\br_{i},\omega)G_0(\bm{r},\br'_{i},\omega)\right], \label{eq:kernel_C}
\end{equation}
where we used seismic reciprocity (the Green's function is unchanged upon exchanging source and receiver).
Equation~\ref{eq:kernel_C} shows that to compute the perturbation to the cross-covariance we need to compute a product of two Green's functions, one with a source at $\br_{i}$ and the other one with a source at $\br'_{i}$.       

\subsection{Travel-Time and Amplitude Sensitivity Kernels}

With the expression in hand for the perturbation to the cross-covariance, we now extract the travel-time and amplitude perturbations from the cross-covariance function.
Using Equation~\ref{eq:travel_time_general} and Equation~\ref{eq:deltaC},  the expectation value of the perturbation to the travel time $[\langle \delta \tau_{i} \rangle]$ and to the amplitude $[\langle \delta a_{i} \rangle]$ is given by 
\begin{equation}
 \langle \delta q_{i} \rangle = 2\pi \int_{-\wn}^{\wn} W_{q}^*(\omega) \langle \delta C_{i}(\omega) \rangle  \mathrm{d}\omega 
= \int_V K_{i}^{q}(\br) \frac{\delta c(\br)}{c_0} \mathrm{d}^3\bm{r}, \quad q \in \{\tau, a\}  ,\label{eq:travel_time_i}
\end{equation}
where $K^{q}$ are the point-to-point  sensitivity kernels
\begin{equation}
K_{i}^{q}(\br) =  2\pi \int_{-\wn}^{\wn} W_{q}^*(\omega) \mathscr{C}_{i}(\br,\omega) \mathrm{d}\omega.  \label{eq:kernel_w}
\end{equation}

Next we need to average the measurements for the deep-focusing technique as explained in Section~\ref{sec:Deep_Foc}. Using Equation~\ref{eq:ave_travel_time}, the expectation values of the averaged travel-time $\langle \overline{\delta \tau} \rangle$ and amplitude $\langle \overline{\delta a} \rangle$  perturbations can be written as
\begin{equation}
 \langle \overline{\delta q}(\br_t) \rangle  = \frac{1}{\Nobs} \sum_{i=1}^{\Nobs} \langle \delta q_{i}(\br_t) \rangle  
 = \int_V \overline{K^{q}}(\bm{r};\br_t) \frac{\delta c(\br)}{c_0} \mathrm{d}^3\bm{r}, \quad q \in \{\tau, a\}   , \label{eq:ave_travel_time_kernel}
\end{equation}
where $\overline{K^q}(\bm{r};\br_t)$ are  the deep-focusing sensitivity kernels targeting a point at $\br_t$ defined by
\begin{equation}
\overline{K^{q}}(\bm{r};\br_t) =  \frac{1}{\Nobs} \sum_{i=1}^{\Nobs}  K_{i}^{q}(\br), \quad q \in \{\tau, a\} .
\label{eq:Ave_k_tau}
\end{equation}

\section{Example Calculations}
\label{sec:results}

\subsection{Choice of Numerical Values and Parameters}
In the following, the value of the reference sound speed is  
$c_0=10^5$~$\rm{m \, s}^{-1}$, the wave  attenuation parameter is $\gamma=10^{-3}$, and   $\mathrm{R_\odot=696}$~Mm. 
The frequency dependence of the source covariance is chosen to be a Gaussian profile,
\begin{equation}
\Pi(\omega) = \exp\left( -\frac{(\left|\omega\right| - \omega_0)^2}{2\sigma^2} \right)  ,\label{eq:Pi_func_Gauss} 
\end{equation}
where $\omega_0 / 2\pi = 3$~mHz  and $\sigma / 2\pi = 1$~mHz. In our computations, we chose a temporal cadence $h_t = 45$~s, \emph{i.e.} the Solar Dynamics Observatory (SDO)/ Helioseismic and Magnetic Imager (HMI) cadence.  

To compute the travel time and the amplitude, we have to define the window function $w$ in  Equations~\ref{eq:weighting_func}\,--\,\ref{eq:weighting_func_amp}. Since in this setup the cross-covariance function has a single branch, we chose a Heaviside step function: 
\begin{equation}
w(t)=
\begin{cases}
1  &\textrm{if} \quad t > 0,\\
0 &\textrm{otherwise}.
\end{cases}
  \label{eq:window_func}
\end{equation} 

\begin{figure}[t]
\begin{center}
\includegraphics[width=0.65\textwidth, trim= {0 0 0 {-0.03\textwidth}}]{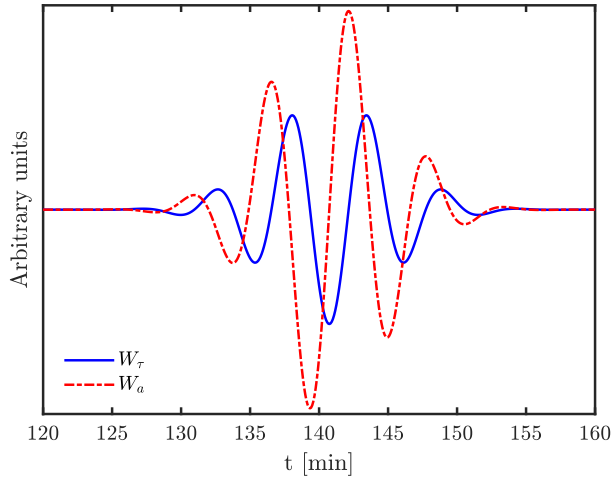}
\caption{Sketch of the weighting functions for measuring travel times $[W_{\tau}]$ and cross-covariance amplitudes $[W_{a}]$. The function $W_a$ is proportional to the unperturbed cross-covariance function $[C_0]$. In this example, the two observation points are separated by a distance $D=1.2\,\mathrm{R_\odot}$. The scalings for the functions  $W_{\tau}$ and $W_{a}$ that are shown here are  arbitrary as the two functions have different units.} 
\label{fig:C_Wtau_Wamp}
\end{center}
\end{figure}

Using the analytic expression for the Green's function (Equation~\ref{eq:analytic_Green}) we obtain the reference cross-covariance $[C_0]$ (Equation~\ref{eq:cross_cov}). Figure~\ref{fig:C_Wtau_Wamp} shows the travel-time and amplitude weighting functions [$W_\tau$ 
and $W_a$] as a function of time for a pair of points separated by a distance of $D=1.2\,\mathrm{R_\odot}$. The function $W_a$ is proportional to $C_0$ as stated by Equation~\ref{eq:weighting_func_amp}, while $W_{\tau}$  is proportional to the temporal derivative of $C_0$ (Equation~\ref{eq:weighting_func}) and is thus shifted by one-fourth of a period.  

\subsection{Point-to-point Sensitivity Kernels}

\begin{figure}[t]
\begin{center}
\includegraphics[width=1\textwidth, trim= {0 0 0 {-0.035\textwidth}}]{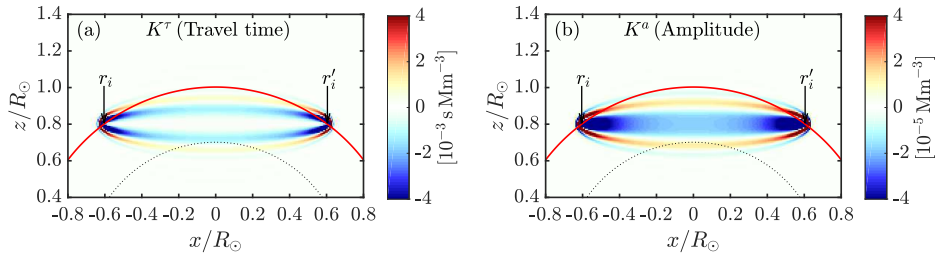}
\caption{2D cross-sections ($y=0$) through the point-to-point kernels for sound-speed perturbations:
({\bf a}) travel-time kernel $K^{\tau}(\br)$ and ({\bf b}) amplitude kernel $K^{a}(\br)$. The pair of points ($\br_i,\br'_i$) on the surface are separated by $D=1.2\,\mathrm{R_\odot}$.  The {\it dotted circle} highlights radius $r=0.7\,\mathrm{R_\odot}$.} 
\label{fig:single_kernels}
\end{center}
\end{figure}

Using Equation~\ref{eq:kernel_w}, we compute the point-to-point travel-time and amplitude sensitivity kernels for sound-speed perturbations with a pair of points separated by $1.2\,\mathrm{R_\odot}$.  Cross-sections through the point-to-point sensitivity kernels for the sound speed are shown in Figure~\ref{fig:single_kernels}. As already discussed in geophysics \citep{Dahlen_Baig_2002} and in helioseismology \citep{Gizon_Birch_2002}, the travel-time kernel $[K^{\tau}]$ has small values along the geometrical ray path and the largest absolute values in the surrounding first Fresnel zone; see  Figure~\ref{fig:single_kernels}(a). The kernel changes sign multiple times away from the ray path when crossing higher-order Fresnel zones. One the other hand, the  amplitude sensitivity kernel for sound-speed takes maximum absolute values along the ray path \citep{Nolet_etal_2005},
see Figure~\ref{fig:single_kernels}(b). 
For a uniform background model, both point-to-point kernels are axially symmetric about the ray path. The total volume integrals of the two-point kernels are negative, $\int K^{\tau}(\br) \mathrm{d}^3\bm{r}\approx -8500$~s and $\int K^{a}(\br) \mathrm{d}^3\bm{r}\approx -1.2$, which means that a uniform reduction in sound speed leads to a longer travel time and a larger cross-covariance amplitude. 

\subsection{Deep-Focusing Sensitivity Kernels}

\begin{figure}[t]
\begin{center}
\includegraphics[width=1\textwidth, trim= {0 0 0 {-0.24\textwidth}}]{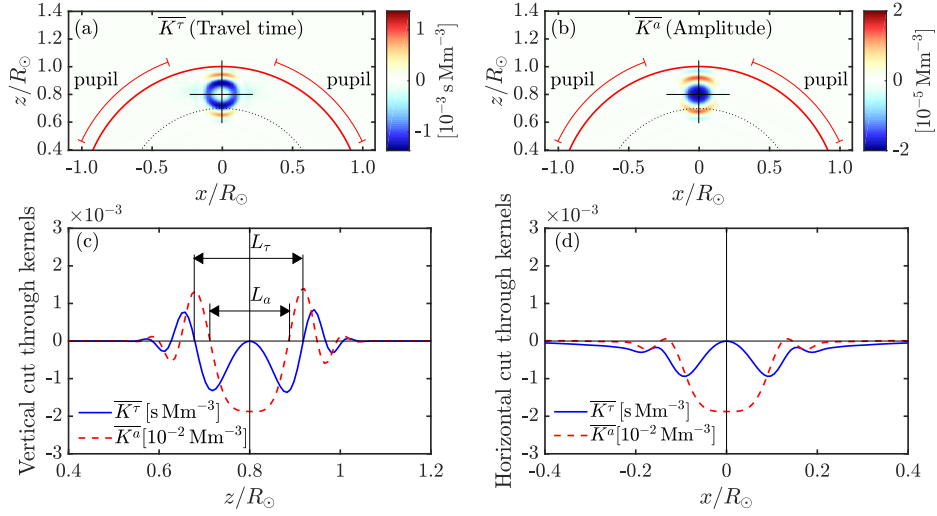}
\caption{2D cross-sections ($y=0$) through the 3D deep-focusing ({\bf a}) sensitivity kernel for $\overline{\delta \tau}$ and ({\bf b}) sensitivity kernel for $\overline{\delta a}$ averaged over the $N=N_{\theta}N_{\phi}$ observation points, where  $N_{\theta}=158$ and $N_{\phi}=793$. The {\it cross} shows the target location, $x_t=y_t=0$, $z_t=0.8\,\mathrm{R_\odot}$ which corresponds to the separation distance, $D=1.2\,\mathrm{R_\odot}$ and the {\it dots} are situated at the surface at $z=0.7\,\mathrm{R_\odot}$. ({\bf c}) {\it Vertical} ($x=0$, $y=0$) and ({\bf d}) {\it horizontal} ($y=0$, $z=0.8\,\mathrm{R_\odot}$) {\it cuts} of the sensitivity kernels. $L_{\tau}$ and $L_{a}$ are the vertical widths of the deep-focusing kernels for travel time and amplitude, respectively.}
\label{fig:aveKernels}
\end{center}
\end{figure}

\begin{figure}[t]
\begin{center}
\includegraphics[width=1\textwidth]{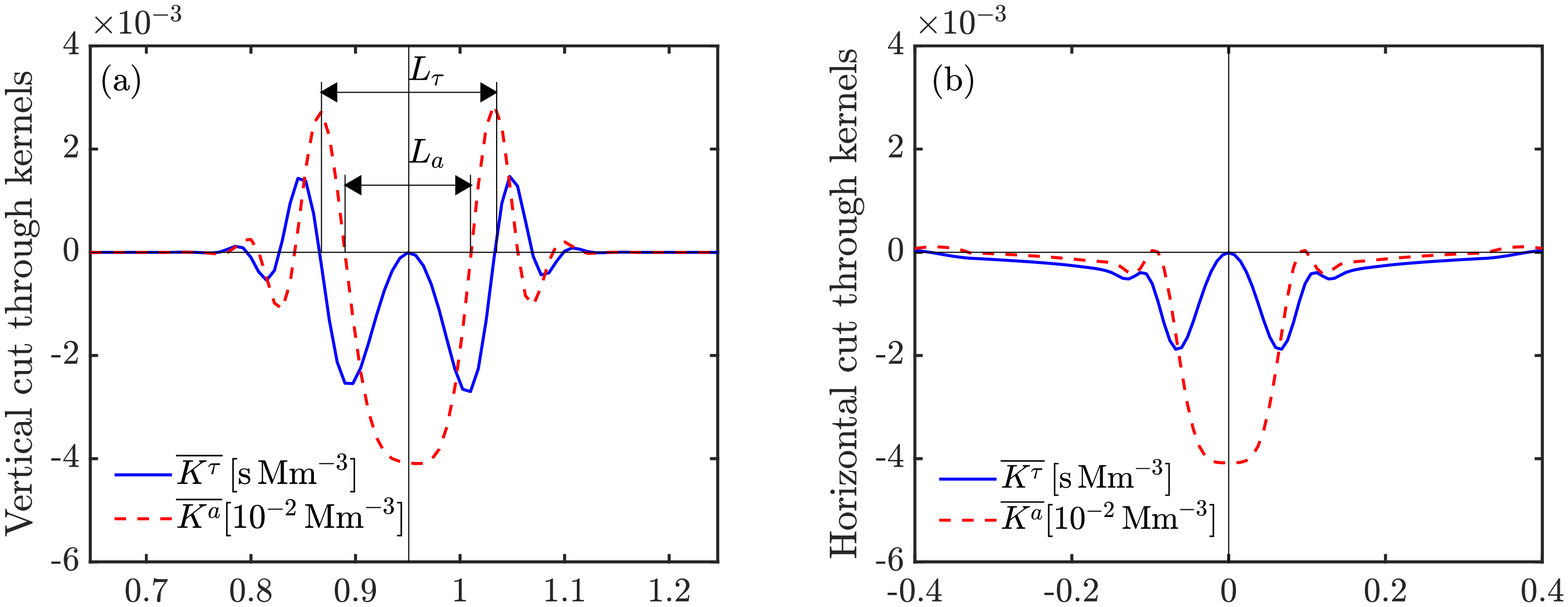}
\caption{({\bf a}) {\it Vertical} ($x=0$, $y=0$) and ({\bf b}) {\it horizontal} ($y=0$, $z=0.95\,\mathrm{R_\odot}$) {\it cuts} of the sensitivity kernels for $\overline{\delta \tau}$ and $\overline{\delta a}$ for the target location, $x_t=y_t=0$, $z_t=0.95\,\mathrm{R_\odot}$ which corresponds to the shorter separation distance, $D=0.63\,\mathrm{R_\odot}$.}
\label{fig:aveKernels_nearSurface}
\end{center}
\end{figure}

With the point-to-point kernels for sound-speed perturbations in hand, we compute the deep-focusing sensitivity kernels for averaged travel time and amplitude using Equation~\ref{eq:Ave_k_tau}. 
We consider all pairs of points in a pupil such that their ray paths intersect at a given target point along the $z$-axis. Neighboring observation points are separated in colatitude by a distance of approximately  $\lambda_{\rm min}/4 \approx 5$~Mm ($0.41^{\circ}$), where $\lambda_{\rm min}$ is the minimum wavelength used in this calculation. For a target point at  radius  $z_t = 0.8\,\mathrm{R_\odot}$, Figures~\ref{fig:aveKernels}(a) and~\ref{fig:aveKernels}(b) show 2D cross-sections ($y=0$) through the deep-focusing sound-speed sensitivity kernels for $\overline{\delta \tau}$ and  $\overline{\delta a}$. For  travel-time measurements, the sensitivity is restricted to a shell surrounding the target location. In the case of  amplitude measurements, the sensitivity is highly localized at the target point. This is a direct consequence of the structure of the point-to-point kernels depicted in Figure~\ref{fig:single_kernels}. Figure~\ref{fig:aveKernels_nearSurface} shows the same bottom panels as in Figure~\ref{fig:aveKernels}, but for a target point near the surface, $z_t=0.95\,\mathrm{R_\odot}$. This target point leads to a shorter separation distance: $D=0.63\,\mathrm{R_\odot}$.

\subsection{Kernel Widths as Functions of Target Depth}

\label{sec.fresnel}

The vertical widths of the deep-focusing sensitivity kernels for travel time [$L_{\tau}$] and amplitude [$L_a$] are defined in Figure~\ref{fig:aveKernels}(c). 
This width indicates the extent of the central regions of a kernel, within which it keeps the same (negative) sign. The smaller  $L_{\tau}$ (or $L_a$), the higher the spatial resolution of the travel-time (or amplitude) deep-focusing technique.

\begin{figure}[t]
\begin{center}
\includegraphics[width=0.7\textwidth]{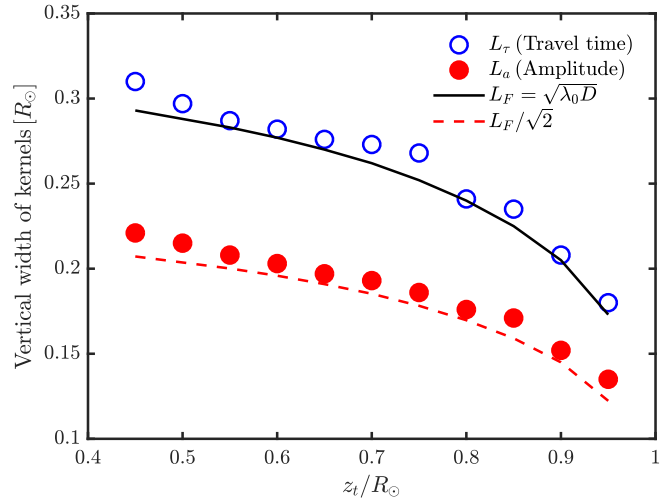}
\caption{Vertical width of the deep-focusing sensitivity kernels for travel time ($L_{\tau}$, {\it open circles}) and amplitude ($L_a$, {\it full circles}) as a function of target position. The {\it solid} and {\it dashed curves} show the simple models described in Section~\ref{sec.fresnel}. The width of the first Fresnel zone is approximately $L_F = \sqrt{\lambda_0 D}$, where $D$ is the distance between the two points and $\lambda_0=33.3$~Mm is the wavelength at frequency $3$~mHz. }
\label{fig:L_width}
\end{center}
\end{figure}

In Figure~\ref{fig:L_width} the widths $L_{\tau}$ and $L_a$ are plotted as functions of the target radius $z_t$. The sensitivity kernels for amplitude measurements  are better localized than those for travel-time measurements, at all depths, with  $L_a \approx 0.7 L_{\tau}$. Furthermore $L_{\tau}$ and $L_a$  increase with target depth.  

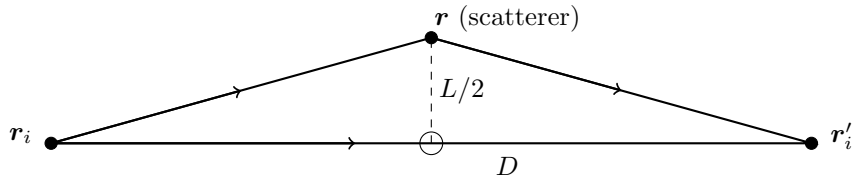
\begin{figure}
\centering
\begin{tikzpicture}
\draw [thick]  (0,0) --(5,1.4) -- (10,0) -- (0,0);
\draw  [->] [thick]  (0,0) --(2.5,0.7);
\draw  [->] [thick] (5,1.4) --(7.5,0.7);
\draw  [->] [thick] (0,0) --(4,0);
\node at (6,-0.3) {$D$};
\draw [thin,dashed]  (5,0) --(5,1.4); \node at (5.4,0.7) {$L/2$};
\draw[fill] (0,0) circle [radius=0.08]; \node at (-0.4,0.1) {$\br_i$};
\draw[fill] (10,0) circle [radius=0.08]; \node at (10.4,0.1) {$\br_i'$};
\draw[fill] (5,1.4) circle [radius=0.08]; \node at (6,1.65) {$\br$ (scatterer)};
\draw (5,0) circle [radius=0.15];
\end{tikzpicture}
\caption{Sketch of the geometry discussed in Section~\ref{sec.fresnel}. A scatterer is placed at equal distance to the two observation points. The scattered and direct wave fields combine at observation point $\br_i'$. The {\it open circle} indicates the position of the target point.}
\label{fig:fresnel} 
\end{figure}

In order to better understand the data points, we consider a simplified version of the model presented in Section~\ref{sec:Born approximation}. For a single sound-speed scatterer (volume $\mathrm{d}V$) at position $\br$ in the mid-plane, the cross-covariance between observation points $\br_i$ and $\br_i'$ can be written
\begin{eqnarray}
\label{eq.phitot}
C(\br_i,\br_i') &\propto& {\rm Im} \left[ G_0(\br_i',\br_i) + \tilde{\epsilon} G_0(\br_i',\br) G_0(\br,\br_i) \right] ,
\quad   \tilde{\epsilon} = 2 k_0^2 \frac{\delta c}{c_0} \mathrm{d}V  ,
\nonumber 
\\
&\propto&   {\rm Im} \left[ G_0(\br_i',\br_i) \left(1 - \epsilon  \, \rm e^{\ii \Delta \phi}  \right) \right] ,
\qquad \qquad \; \epsilon 
\approx \frac{\tilde{\epsilon}}{\pi D}. 
\end{eqnarray}
In the above expression, $|\epsilon|\ll 1$ is the scattering amplitude. The phase perturbation [$\Delta\phi$] is due to the difference  [$\delta$] between the path through the scatterer and the direct path between the two points [$D$], 
\begin{equation}
\Delta \phi = 2\pi \frac{ \delta }{ \lambda_0 } ,
\end{equation}
where $\lambda_0=33.3$~Mm is the  reference wavelength. For a scatterer at equal distance from the two points (see Figure~\ref{fig:fresnel}), we have
\begin{equation}
\delta = 2 \sqrt{(D/2)^2 + (L/2)^2} - D \approx \frac{L^2}{2D}  .
\end{equation}
Thus the phase perturbation due to a scatterer midway between the two points and at a distance $L/2$ from the direct path is approximately 
\begin{equation}
\Delta \phi \approx \frac{\pi L^2}{\lambda_0 D} .
\end{equation}

A travel time $[\tau]$  is most easily interpreted as a phase travel time. 
According to Equation~\ref{eq.phitot}, there is no phase change at the receiver when ${\rm Arg} (1 - \epsilon  \rm e^{\ii \Delta \phi} ) = 0$, \emph{i.e.} when $\Delta \phi = n \pi$,  $n\in \mathbb{Z}$. In particular the sensitivity is zero along the direct ray path ($n=0$). The width of the travel-time kernel coincides with the boundary of the first ($n=1$) Fresnel zone:
\begin{equation}
\Delta \phi_\tau = \pi \quad \text{ and }  
\quad L_\tau= \sqrt{\lambda_0 D} .
\label{eq.Ltau}
\end{equation}
This simple result was reported earlier in 2D by \citet{Gizon_2006}.

Contrary to the travel-time perturbations, cross-covariance amplitude perturbations are extremal along the ray path (where $\Delta\phi=0$). The amplitude of the cross-covariance is unchanged when $\left| 1 -  \epsilon \, \rm e^{\ii\Delta \phi} \right| = 1$. For small-amplitude perturbations, this condition is approximately  $2 \, \epsilon \cos \, \Delta \phi = 0$, \emph{i.e.} $\Delta \phi = \pi/2 + m \pi$, with $m\in \mathbb{Z}$. Setting $m=0$ gives the width of the amplitude kernel:
\begin{equation}
\Delta \phi_a = \pi/2
\quad \text{ and }  
\quad L_a = \sqrt{\lambda_0 D/2} .
\label{eq.La}
\end{equation}

The dependence of the widths on target radius $z_t$ is understood through $D = 4 \sqrt{R^2- z_t^2}$, so that $L\propto (R^2 - z_t^2)^{1/4}$. As seen in Figure~\ref{fig:L_width}, the model values for $L_\tau$ and $L_a$ from Equations~\ref{eq.Ltau}\,--\,\ref{eq.La} are in reasonable agreement with the numerical values, including the relationship $L_a =  L_\tau/\sqrt{2}\approx 0.7 L_\tau$.


\subsection{Noise Covariance}

\begin{figure}[t]
\begin{center}
\includegraphics[width=0.6\textwidth, trim= {0 0 0 {-0.03\textwidth}}]{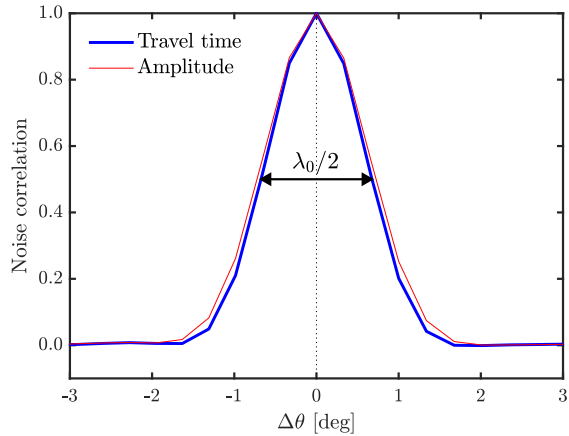}
\caption{Cut through the noise correlation matrix for point-to-point travel times ({\it solid curve}) and point-to-point amplitude measurements ({\it thin-red curve}) as a function of the angular distance between neighboring points. 
The reference observation point is kept fixed at $\theta = 45.2^{\circ}$ and the target radius is  $z_t=0.8\,\mathrm{R_\odot}$. 
The {\it double-headed arrow} indicates  $\lambda_0/2$, where $\lambda_0=33.3$ Mm is the wavelength at frequency $3$~mHz.}
\label{fig:noise_corr}
\end{center}
\end{figure} 

The model for the noise covariance matrix for travel-time and amplitude measurements was outlined in Section~\ref{sec:noise_model}. Figure~\ref{fig:noise_corr} shows a cut through the noise correlation matrix of point-to-point travel times and amplitudes. The correlation between neighboring pairs of points drops fast as  a function of angular distance. For both travel-time and amplitude measurements, the correlation distance at half maximum is approximately $\lambda_0/2$  \citep[see also][]{Gizon_Birch_2004}. This justifies a \emph{posteriori} why we chose  points in the pupil that are separated by $\lambda_0/4\approx 8.3$~Mm to avoid under sampling.

\subsection{Localized Sound-Speed Anomaly at $z_0=0.7\,\mathrm{R_\odot}$}

\begin{figure}[t]
\begin{center}
\includegraphics[width=1\textwidth, trim= {0 0 0 {-0.15\textwidth}}]{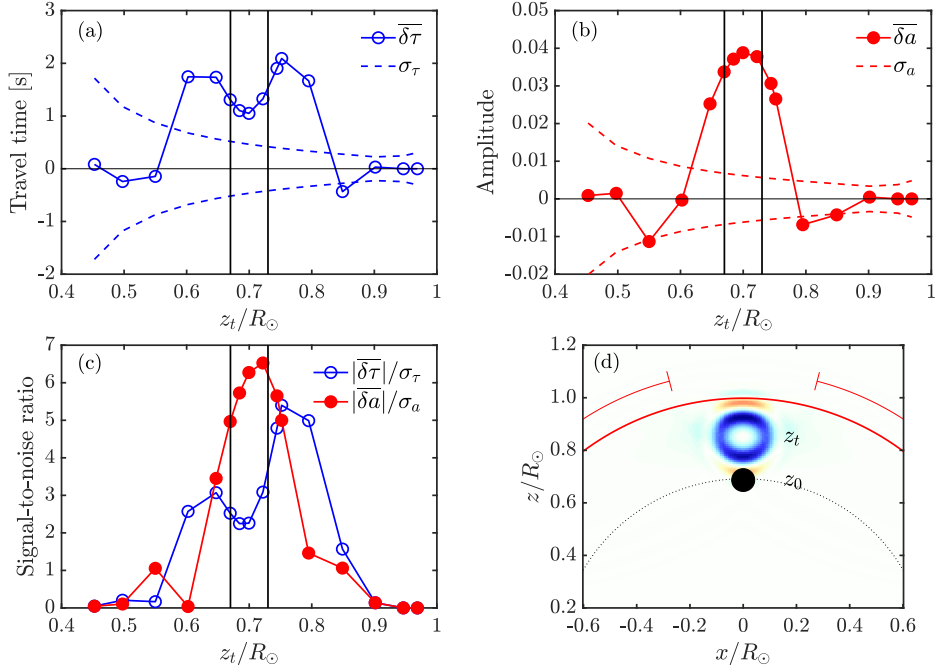}
\caption{({\bf a}) Deep-focusing travel-time perturbation $\overline{\delta \tau}$ (connected {\it open circles}) and ({\bf b}) amplitude perturbation $\overline{\delta a}$ (connected {\it filled circles})  due to 2\,\% decrease in sound speed at radius $z_0=0.7\,\mathrm{R_\odot}$ given by Equation~\ref{eq:target_func1}. The {\it dashed lines} show the 1-$\sigma$ level of stochastic noise for an observation duration $T=$~four years. The {\it vertical lines} indicate the width of the negative sound-speed perturbation ($z_0 \pm s$). 
({\bf c}) Signal-to-noise ratios for the deep-focusing travel-time and amplitude perturbations as functions of target radius. ({\bf d}) Geometrical setup. The (negative) Gaussian perturbation in sound speed is indicated by the {\it filled black circle} at $\br_0=(x_0,y_0,z_0)=(0,0,0.7\,\mathrm{R_\odot})$. An example deep-focusing sensitivity kernel for $\overline{\delta \tau}$ is shown for the target radius $z_t=0.85\,\mathrm{R_\odot}$.}
\label{fig:d_tau_amp_pupil_target1}
\end{center}
\end{figure} 

In order to quantify the bias and variance of the travel-time and amplitude measurements in the present deep-focusing setup, we compute the travel-time and amplitude perturbations generated by sound-speed perturbations of our choosing (forward modeling).

In this section we consider a highly localized perturbation in sound speed with a Gaussian profile centered at $\br_0=(x_0,y_0,z_0)=(0,0,0.7\,\mathrm{R_\odot})$ such that 
\begin{equation}
\frac{\delta c(\br)}{c_0} = -A_1 \exp \left( -\|\br - \br_0 \|^2/2s^2 \right), 
\label{eq:target_func1}
\end{equation}
where  $A_1=0.02$. Notice that we have chosen a negative perturbation in sound speed. The parameter $s = 0.03\,\mathrm{R_\odot}$ determines the extent of the perturbation, which is roughly of the same size as the wavelength ($\lambda_0\approx 0.05\,\mathrm{R_\odot}$). The location of the perturbation is represented by the filled black circle in Figure~\ref{fig:d_tau_amp_pupil_target1}(d).

Figure \ref{fig:d_tau_amp_pupil_target1}(a) shows the deep-focusing travel-time measurements $[\overline{\delta \tau}]$ and the corresponding noise levels (standard deviations) for different target locations $\br_t = (0,0,z_t)$, where $0.4<z_t/\mathrm{R_\odot}<1$. Due to the hollow nature of the deep-focusing travel-time kernel, the signal is weaker at the depth where the perturbation is located than in the surroundings. The bulk of the perturbation is within $| z_0-z_t | < L_\tau(z_0) /2 \approx 0.13\,\mathrm{R_\odot}$. On the other hand, a  maximum signal for the amplitude measurements is obtained at the  radius where the perturbation is placed   (Figure \ref{fig:d_tau_amp_pupil_target1}(b))
 due to the concentrated sensitivity of the deep-focusing kernel for amplitude measurements (Figure \ref{fig:aveKernels}(b)). 
 
To compare the two types of measurements, travel-time \emph{versus} amplitude measurements, the signal-to-noise ratios are plotted in Figure~\ref{fig:d_tau_amp_pupil_target1}(c). We find that the signal-to-noise ratio is higher and better localized for the amplitude measurements than for the travel-time measurements, given the highly localized perturbation in sound speed that we chose in this section.
 
\subsection{Sound-Speed Anomaly in a Shell at Radius $r_0=0.7\,\mathrm{R_\odot}$}

\begin{figure}[t]
\begin{center}
\includegraphics[width=1\textwidth, trim= {0 0 0 {-0.15\textwidth}}]{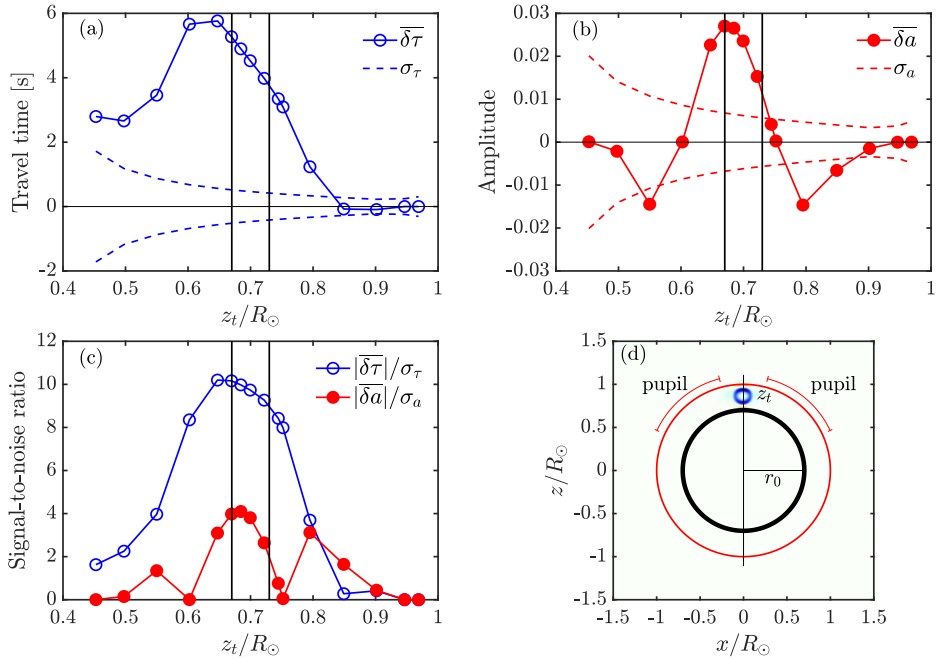}
\caption{({\bf a}) Deep-focusing travel-time perturbation $\overline{\delta \tau}$ and ({\bf b}) amplitude perturbation $\overline{\delta a}$ due to a $0.2$\,\% decrease in sound speed in a thin shell at radius $r_0=0.7 \,\mathrm{R_\odot}$ and defined by Equation~\ref{eq:target_func2}. 
The {\it vertical lines} indicate the width of the negative sound-speed perturbation ($r_0 \pm s$). 
({\bf c}) Signal-to-noise ratios for the deep-focusing  travel-time and amplitude perturbations {\it versus} target depth, for an observation duration $T=$~four years. ({\bf d}) 
The {\it thick black circle} of thickness $2s = 0.06\,\mathrm{R_\odot}$ indicates the location of the shell of sound-speed perturbation. The {\it blue shades} show a cut through a deep-focusing travel-time sensitivity kernel with target radius $z_t=0.85\,\mathrm{R_\odot}$.}
\label{fig:d_tau_amp_pupil_target2}
\end{center}
\end{figure} 

The search for solar-cycle changes at the bottom of the solar convection zone is a key question in helioseismology. In this section we consider a shell of perturbation in sound speed at radius $r_0 = 0.7\,\mathrm{R_\odot}$ with a profile given by 
\begin{equation}
\frac{\delta c(\br)}{c_0} = -A_2\exp \left( -\left( \| \br \| - r_0 \right)^2/2s^2 \right), 
\label{eq:target_func2}
\end{equation}
where $A_2 = 0.002$ and $s = 0.03\,\mathrm{R_\odot}$. This shell of negative sound-speed perturbation is illustrated in Figure \ref{fig:d_tau_amp_pupil_target2}(d). As in the previous section the radial extent of this perturbation is of the order of a wavelength.

The corresponding travel-time and amplitude perturbations, as well as  the noise levels for $T=$~four years, are shown in Figures \ref{fig:d_tau_amp_pupil_target2}(a) and \ref{fig:d_tau_amp_pupil_target2}(b). We see that the travel-time and amplitude signals peak below $z_t=0.7\,\mathrm{R_\odot}$: the  deep-focusing averaging scheme is not unbiased. For a shell-like perturbation, the signal-to-noise ratio is twice as large for the travel-time measurements as for the amplitude measurements (Figure \ref{fig:d_tau_amp_pupil_target2}(c)).

\section{Conclusion}
\label{sec:summary}
In this article we considered   toy models in a uniform background medium  to  study the localization and noise properties of the deep-focusing time--distance technique. We considered  two measurement quantities extracted from the cross-covariance function:  travel times and amplitudes. The sensitivity kernels for sound speed were derived under the first Born approximation.

We computed the spatial sensitivity of travel-time and amplitude to perturbations in sound speed with respect to a uniform background medium. We find that the travel-time sensitivity to sound-speed perturbations is zero at the target location and negative in a surrounding region with diameter $L_{\tau} \approx (\lambda_0 D)^{1/2}$, where $\lambda_0$ is the wavelength and $D$ is the travel distance between the points used in the deep-focusing averaging. On the other hand, the amplitude sensitivity peaks at the target location and is negative in a region with diameter $L_a \approx (\lambda_0 D/2)^{1/2}$, resulting in a higher signal-to-noise ratio for small-scale perturbations. We conclude that amplitude measurements are a useful complement to  travel-time measurements  in local helioseismology. 

In future studies, we intend to extend this work to a standard solar model using accurate computations of Green's functions in the frequency domain. We also intend to study the capability of the deep-focusing technique to recover flows in the solar interior.
Deep-focusing travel times have already been used to recover  meridional circulation  \citep[\emph{e.g.}][]{Rajaguru_Antia_2015}. No significant improvement is expected from using deep-focusing amplitude measurements to recover such slowly varying flows. However, amplitude measurements should help resolve flows that vary on scales smaller than the wavelength, \emph{e.g.} convective flows.

\begin{acks}
We thank Thomas~L. Duvall, Aaron~C. Birch, Zhi-Chao Liang, Chris~S. Hanson and Kaori Nagashima for useful discussions and comments. MP is a member of the International Max Planck Research School (IMPRS) for Solar System Science at the University of G\"ottingen.
\end{acks}


\bibliographystyle{spr-mp-sola}
\bibliography{ms.bib}  

\end{article} 
\end{document}